\documentclass[preprint,12pt]{elsarticle}

\usepackage{amssymb}
\usepackage{amsmath}
\usepackage{bm}

\usepackage{siunitx}
\usepackage{upgreek}
\usepackage[version=4]{mhchem}
\usepackage{makecell}
\usepackage{multirow}

\usepackage[linesnumbered,lined,boxed,commentsnumbered]{algorithm2e}

\usepackage{enumitem}
\usepackage{subcaption}

\usepackage{hyperref}

\usepackage{nomencl}
\makenomenclature
\renewcommand\nomgroup[1]{%
  \item[\emph{
  \ifstrequal{#1}{A}{Acronyms}{%
  \ifstrequal{#1}{B}{Roman symbols}{%
  \ifstrequal{#1}{C}{Greek symbols}{%
  \ifstrequal{#1}{E}{Superscripts}{
  \ifstrequal{#1}{D}{Subscripts}{
  \ifstrequal{#1}{O}{Other symbols}{%
  }}}}}}%
}]}

\newcommand{\nomRoman}[1][]{\nomenclature[B,#1]}
\newcommand{\nomGreek}[1][]{\nomenclature[C,#1]}

\newcommand{\nomSub}[1][]{\nomenclature[D,#1]}
\newcommand{\nomAcro}[1][]{\nomenclature[A,#1]}

\journal{Journal of Energy Storage}

\begin{document}

\begin{frontmatter}



\title{Packed bed thermal energy storage for waste heat recovery in the iron and steel industry: An experimental study on powder hold-up and pressure drop}


\author[IET]{Paul Schwarzmayr\corref{corrauthor}}\ead{paul.schwarzmayr@tuwien.ac.at}
\author[IET]{Felix Birkelbach}
\author[IET]{Heimo Walter}
\author[VASD]{Florian Javernik}
\author[VASD]{Michael Schwaiger}\ead{michael.schwaiger@voestalpine.com}
\author[IET]{René Hofmann}\ead{rene.hofmann@tuwien.ac.at}

\cortext[corrauthor]{Corresponding author}

\affiliation[IET]{organization={Institute for Energy Systems and Thermodynamics, TU Wien},
            addressline={Getreidemarkt 9}, 
            city={Vienna},
            postcode={1060}, 
            country={Austria}}
\affiliation[VASD]{organization={voestalpine Stahl Donawitz GmbH},
            addressline={Kerpelystraße 199}, 
            city={Leoben},
            postcode={8700}, 
            country={Austria}}



\begin{abstract}
Waste heat recovery in the energy intensive industry is one of the most important measures for the mitigation of climate change. The utilization of just a fraction of the theoretically available waste heat potential would lead to a significant reduction of the primary energy consumption and hence a reduction of greenhouse gas emissions. The present study examines the integration of a packed bed thermal energy storage for waste heat recovery in the iron and steel industry. Along with the highly fluctuating availability of excess heat the main difficulty of waste heat recovery in industrial processes is the high amount of powder that is transported by the hot exhaust gases. Therefore, the experimental investigations in this study focus on the powder hold-up and pressure drop in a packed bed thermal energy storage that is operated with a gas-powder two phase exhaust gas as heat transfer fluid. The ultimate goal is, to assess its suitability and robustness under such challenging operational conditions. The results indicate, that 98 \% of the powder that is introduced into the system with the heat transfer fluid during charging accumulates in the packed bed. Remarkably, most of the powder hold-up in the packed bed is concentrated near the surface at which the heat transfer fluid enters the packed bed. When reversing the flow direction of the heat transfer fluid to discharge the storage with a clean single phase gas, this gas is not contaminated with the powder that has been accumulated in previous charging periods. Furthermore, the radial distribution of the powder hold-up in the packed bed is observed to be even which indicates that there is no risk of random flow channel formation that could affect the thermal performance (storage capacity, thermal power rate) of the system. The entirety of these findings reinforces the great potential of packed bed thermal energy storage systems for waste heat recovery in the energy intensive industry. 
\end{abstract}


\begin{highlights}
\item Thermal energy storage for waste heat recovery in the iron and steel industry.
\item Direct use of exhaust gas (gas-powder two phase flow) as heat transfer fluid.
\item Experimental investigation of powder hold-up and pressure drop.
\item During charging powder accumulates at the surface where it enters the packed bed.
\item During discharging clean heat transfer fluid is not contaminated with powder.

\end{highlights}


\end{frontmatter}



\section{Introduction}
\label{sec:intro}
The waste heat potential from the industry sector is enormous and its exploitation can lead to substantial primary energy savings. Bianchi et al. \cite{bianchi_estimating_2019} estimated the theoretical waste heat potential in the European Union (EU) industry in 2014 to be \SI{918}{\tera\watt\hour}. This is nearly 8 \% of the EU's annual final energy consumption and its utilization as a substitute for heat generated from burning natural gas would avoid a significant amount
of $\mathrm{CO_2}$ that is emitted into the atmosphere per year. The biggest challenge of waste heat recovery is, that the utilization of only a fraction of the theoretical waste heat potential (approx. one third in 2014) is economically feasible. This is mainly because of techno-economic constraints like minimum temperature requirements, discontinuous waste heat availability, technology costs, or even the lack of suitable technologies. In the iron and steel industry, which is responsible for 18 \% of the EU industry's final energy consumption, excess heat is often unused because of the temporal mismatch between heat availability and heat demand and because of the lack of suitable technologies that are able to operate reliably even under harsh conditions. A system that is to be used for waste heat recovery in the iron and steel industry has to  handle high temperatures and discontinuities in waste heat availability, it has to be energy- and cost efficient as well as robust against challenging operational conditions.

A packed bed thermal energy storage (PBTES) is a type of thermal energy storage (TES) that meets most of these requirements. A comprehensive review considering the implementation of TES systems for industrial waste heat recovery is provided by Miró et al. \cite{miro_thermal_2016}. Since PBTES systems use a non-pressurized steel vessel as storage tank, rocks or some other type of solids as storage material and a gaseous medium as heat transfer fluid (HTF) they are extremely cost efficient and require little maintenance. 
High power rates can be realized even at small temperature differences compared to conventional heat exchangers as the HTF gets in direct contact with the storage material, which leads to an enhanced heat transfer between HTF and storage material \cite{Kalantari_analysis_2022}. Due to the direct contact heat transfer PBTES systems are also robust against erosion, abrasion and fouling. They therefore are especially suitable for waste heat recovery in scenarios where conventional heat recovery systems (heat exchangers) would exceed their technological limits and deteriorate quickly.

Thanks to a huge amount of research during recent years, the thermal behaviour of PBTES systems is understood well. The most important key performance indicators of energy storage systems are energy and exergy efficiency. For PBTES systems, they are determined by various factors such as the geometries of the storage and the storage material and the HTF mass flow rate. Marti et al. \cite{marti_constrained_2018} and Trevisan et al. \cite{trevisan_packed_2021} conducted studies where they used multi-objective optimization techniques to find optimal storage parameters (tank diameter/height ratio, storage material particle size, HTF mass flow rate, ...) with the objective to minimize investment costs and to maximize exergy efficiency. In a well designed PBTES an effect called temperature stratification is utilized to maximize its exergy efficiency. This means, that for a partially charged storage, the storage volume is separated into a hot and a cold zone. The thin volume slice that separates these two zones is called thermocline and is characteristic for the exergy efficiency of the storage. Detailed studies on effects like thermocline degradation in PBTES systems are provided by multiple authors. 
Al Azawii et al. \cite{al-azawii_thermocline_2023} investigated the thermal behaviour of a commercial-scale PBTES under repeated charge/discharge cycles, focusing on total exergy efficiency. Schwarzmayr et al. \cite{schwarzmayr_standby_2023} experimentally examined thermocline degradation and exergy efficiency of a PBTES in standby mode for different HTF flow directions on a lab-scale test rig. For a detailed review on the design of PBTES systems, the selection of suitable storage materials and HTFs, pressure loss and economic aspects of PBTES systems the authors refer to the studies from Esence et al. \cite{esence_review_2017} and Gautam et al. \cite{gautam_review_2020, gautam_review_2020-1}.

\begin{figure}
    \centering
    \includegraphics[width=9cm]{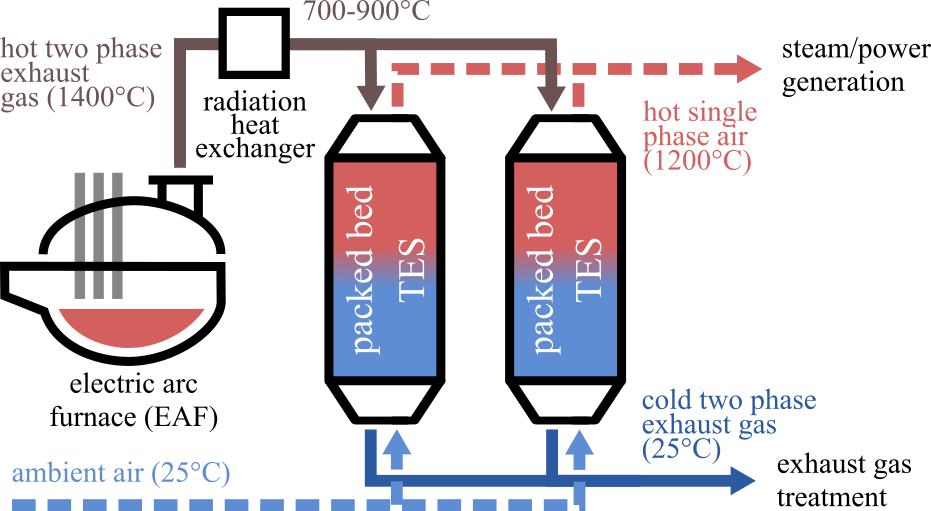}
    \caption{Integration of PBTES systems for the waste heat recovery from EAF exhaust gas}
    \label{fig:usecase}
\end{figure}

Contrary to the maturity of PBTES systems for their integration into concentrated solar power (CSP) plants \cite{geissbuhler_assessment_2019, geissbuhler_assessment_2019-1} studies on their utilization as industrial waste heat recovery systems are rare. 
Ortega-Fernández et al. \cite{ortega-fernandez_thermal_2019} investigated the efficiency of two parallel vertical PBTES systems for waste heat recovery in a steel production plant from the exhaust gases of an electric arc furnace (EAF). A single horizontal PBTES system was proposed and tested by Slimani et al. \cite{slimani_horizontal_2023} for a similar (EAF) use-case. Both studies conclude that the utilization of PBTES systems for waste heat recovery in industrial energy systems leads to significant energy-, cost- and $\mathrm{CO_2}$-emission savings. But yet, the biggest difference between integrating a PBTES into an industrial waste heat recovery system and into a CSP plant is the composition of the HTF. In a CSP plant a PBTES system is charged and discharged with clean air whereas in an industrial waste heat recovery system the HTF will be some kind of gas-powder two phase exhaust gas. The above mentioned studies \cite{ortega-fernandez_thermal_2019, slimani_horizontal_2023} dealt with this issue by placing a high temperature dust filter and a gas-to-gas heat exchanger between the industrial waste heat source and the TES so that the TES can be operated with clean air. However, this approach is far from optimal, because the filtration of high temperature gas is difficult and expensive and the exergy efficiency of gas-to-gas heat exchangers is low. Additionally the lifetime of these heat exchangers would be short due to the abrasiveness of the gas-powder two phase exhaust gas. Therefore, the authors of the present study consider the direct use of high temperature gas-powder two phase exhaust gas from industrial processes as HTF for charging a PBTES. A schematic view of the proposed waste heat recovery system is depicted in Figure \ref{fig:usecase}.
This approach drastically decreases investment (no high temperature filtration, no additional heat exchanger) and maintenance costs (filter, heat exchanger) of the whole system. In order to make assessments on the suitability and robustness of a PBTES system in such a setting this study examines the behaviour of a PBTES system that is operated with a gas-powder two phase exhaust gas as HTF.

The first studies that consider gas-powder two phase flows in packed beds date back to the early 1990s and were conducted to investigate the behaviour of coal powder that is injected into a blast furnace. In the most recent study from Gupta et al. \cite{gupta_quantitative_2022} the authors state that despite the variety of publications considering this topic \cite{zhou_numerical_2021, dong_gaspowder_2004, dong_gaspowder_2004-1, takahashi_permeation_2011, kiochiro_pressure_1991}, the literature reveals that consistency is lacking and that the behaviour of gas-powder two phase flow in packed beds is still not fully understood yet. Additionally, the application of experimental results that were generated in the context of pulverized coal injection into a blast furnace to a PBTES that is operated with a gas-powder two phase HTF is impracticable. Operational conditions that prevail in a blast furnace are fundamentally different to the conditions in a PBTES. Probably the biggest difference is, that in a blast furnace the powder (pulverized coal) undergoes a chemical reaction either with the gaseous part of the flowing fluid (burning) or the packed bed particles (reduction of iron ore) whereas in a PBTES the interactions between the packed bed and the HTF are limited to heat transfer, momentum transfer and adhesion. In a blast furnace coal powder with a narrow particle size distribution and a median particle size of $\SI{75}{\upmu\meter}$ is laterally injected into the bottom of the packed bed. In a PBTES that is operated with a gas-powder two phase HTF the gas-powder flow enters the packed bed through the top surface. Furthermore, the particle size distribution of metal dust from a steel producing processes is much wider with a median particle size of less than $\SI{10}{\upmu\meter}$. Therefore, the authors of this study decide to build upon existing research on gas-powder two phase flows in packed beds that was conducted in the context of coal powder injection into blast furnaces and to apply and extend this area of research towards PBTES systems that are operated with a gas-powder two phase HTF.

The remainder of this paper includes a presentation of the test rig that is designed and erected for the experimental investigations in Section \ref{sec:methods}. Additionally, properties and other information about the materials that are used for the packed bed and the powder are summarized in Section \ref{sec:methods}. Section \ref{sec:analysis} delineates the data analysis procedure and empirical pressure drop equations that are used for measurement data validation. Section \ref{sec:results} includes the presentation of all the results from the experiments as well as an interpretation of these results with the ultimate goal to assess the suitability of PBTES systems for waste heat recovery in the iron and steel industry.


\section{Material and Methods}
\label{sec:methods}
\subsection{Experimental setup}
\label{sec:setup}
\begin{figure}
    \centering
    \includegraphics[width=9cm]{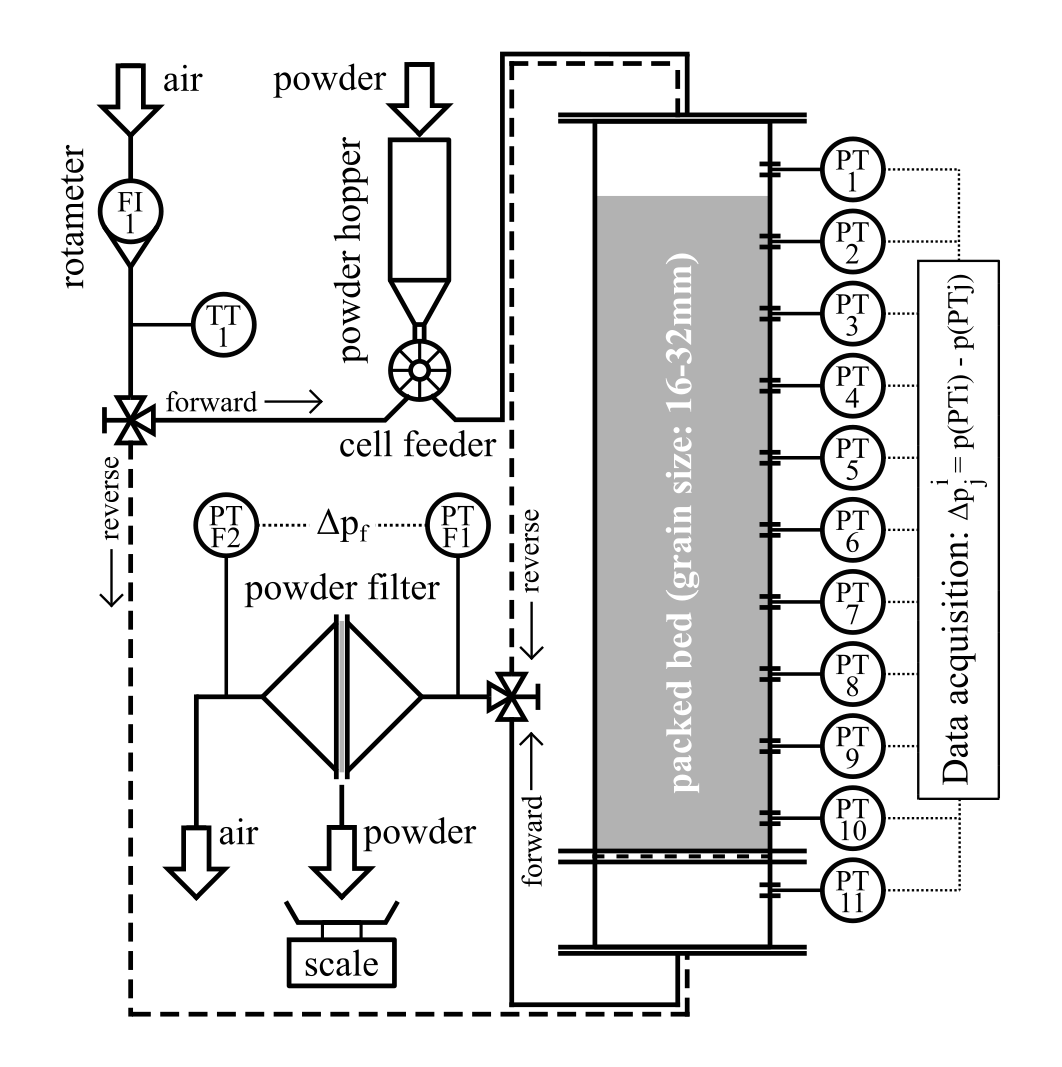}
    \caption{Experimental test rig}
    \label{fig:testrig}
\end{figure}

To investigate the powder hold-up and pressure drop in a PBTES when it is operated with a gas-powder two phase HTF, a lab-scale test rig of a vertical PBTES is used. The geometry of the storage tank, the storage material and the powder for the experiments were chosen in a way, that the operational conditions are comparable with an industrial scale PBTES. Figure \ref{fig:testrig} shows a P\&ID of the test rig with all its components and instrumentation. The storage tank itself is a vertical acrylic glass cylinder with a height to diameter ratio of approximately $3$ and is filled with $\SI{68.5}{\kilo\gram}$ of storage material. As storage material slag, a by-product from the iron and steel industry, is used. In addition to the extremely low costs, the suitability of slag as storage material for a PBTES is justified by its exceptional heat transfer properties due to its geometric shape. The irregular shaped and partly porous rocks that the slag is composed of lead to a uniform random packing, hence an even perfusion, and an improved heat transfer between HTF and storage material. More details about the storage tank's geometry and properties of the storage material are summarized in Table \ref{tab:param}. For the experiments the storage tank is equipped with $11$ pressure measuring points (PT1, PT2, ..., PT11) that are evenly distributed over the height of the packed bed. Piezoresistive pressure sensors are used to record the pressure differences between each of the pressure measuring points in the storage tank. Before the experiments the piezoresistive sensors were calibrated to an accuracy of $\pm 0.06 \%$ of full scale. The flow rate of the HTF (dry, clean ambient air provided by an air supply unit) into the system is controlled with a rotameter flow meter.

\begin{figure}
    \centering
    \includegraphics[width=9cm]{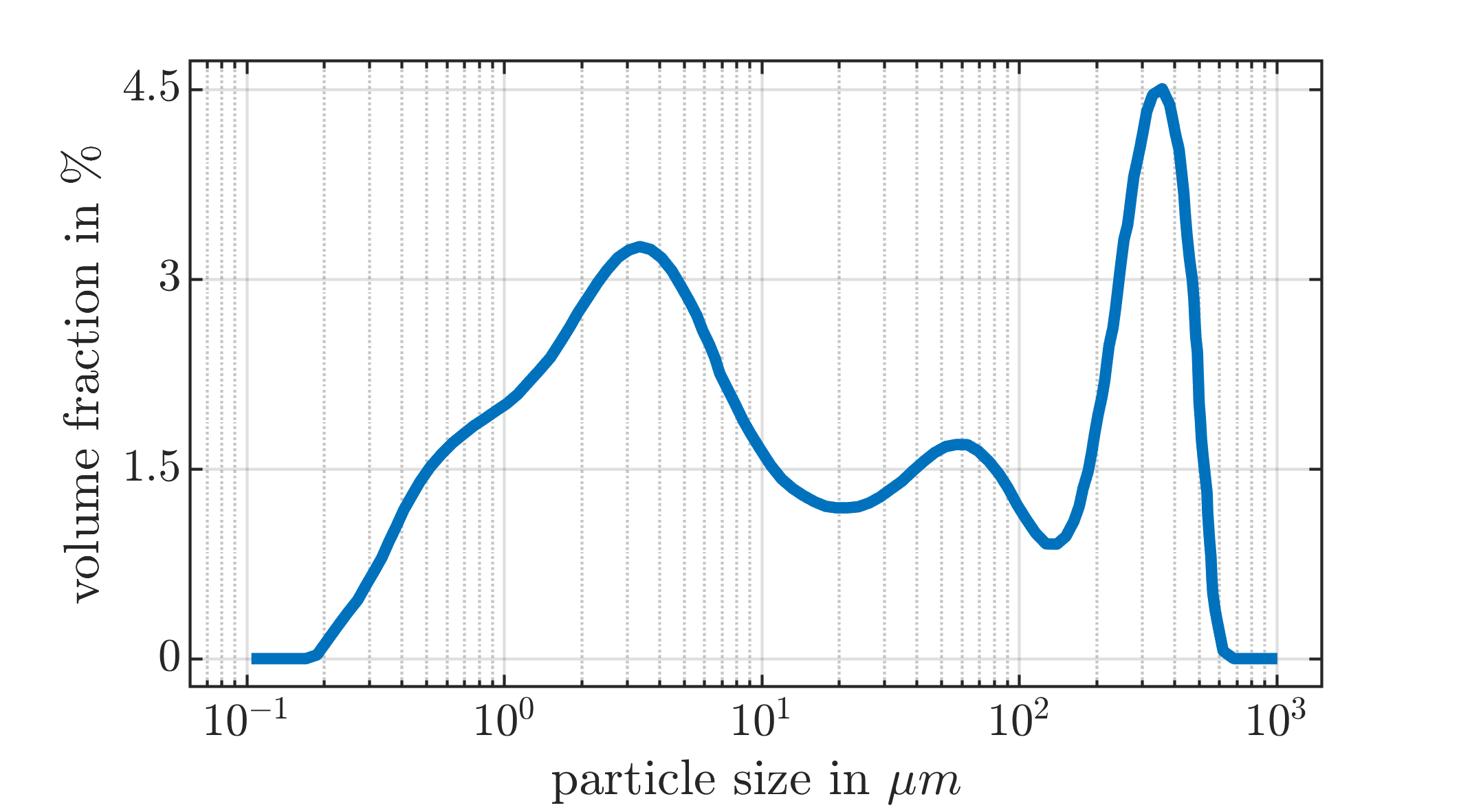}
    \caption{Particle size distribution of the powder used for the experiments.}
    \label{fig:psd}
\end{figure}

To simulate a charging process of the PBTES with a gas-powder two phase flow the initially clean HTF is directed to a cell feeder where powder is added before it enters the PBTES storage tank. The feed rate of the powder can be controlled by adjusting the rotational speed of the cell feeder. In order to represent reality as accurately as possible metal dust collected from the exhaust gases of a steel producing process is utilized as powder for the experiments. In Figure \ref{fig:psd}, the particle size distribution of the powder is provided. The particle size of this type of powder ranges from $\SIrange{0.2}{600}{\mathrm{\upmu}\meter}$ with a median of $\SI{8.85}{\upmu\meter}$ where the most common particle sizes are $\SI{3.5}{\upmu\meter}$ and $\SI{350}{\upmu\meter}$. Detailed information about the metal dust are provided in Table \ref{tab:param}. Downstream of the cell feeder the gas-powder two phase HTF enters the storage tank from the top, passes through the packed bed and leaves the tank at the bottom. Before the HTF exits the system it passes through a powder filter that seperates the remaining powder from the HTF flow. The path of the HTF flow for a charging process is indicated with solid lines in Figure \ref{fig:testrig}. 

For the simulation of a discharging process of the PBTES with clean single-phase HTF, air from the rotameter directly enters the storage tank from the bottom, passes through the packed bed and leaves the tank at the top. Again, the air that exits the storage tank is directed through the powder filter before it is released into the environment. The HTF flow path for the discharging process is indicated with the dashed lines in Figure \ref{fig:testrig}.

\begin{table}
\begin{center}
\caption{Summary of parameters: Test rig geometry, data/properties of storage material, operational parameters}
\label{tab:param}
\begin{tabular}{l l} 
\hline
\multicolumn{2}{l}{\textbf{Test rig}}\\ \hline
Storage type                    & vertical PBTES\\
Tank diameter         & \SI{238}{\milli\meter}\\
Tank height         & \SI{780}{\milli\meter}\\
Tank volume         & \SI{0.034}{\cubic\meter}\\
Packed bed height        & \SI{700}{\milli\meter}\\
Packed bed volume         & \SI{0.031}{\cubic\meter}\\
Packed bed mass             & \SI{68.5}{\kilo\gram}\vspace{0.3cm}\\

\hline
\multicolumn{2}{l}{\textbf{Packed bed material}}\\ 
\hline
\makecell[l]{Type of material\\}           & \makecell[l]{slag (irregular\\ shaped, partly porous)} \\
Particle size                               & \SIrange{16}{32}{\milli\meter}\\
Particle density          & \SI{3800}{\kilo\gram\per\cubic\meter}\\
Void fraction                & 0.42\\
Bulk density          & \SI{2200}{\kilo\gram\per\cubic\meter}\vspace{0.3cm}\\\hline 
\multicolumn{2}{l}{\textbf{Powder}}\\ \hline
\makecell[l]{Type of material\\}           & \makecell[l]{metal dust from the\\ iron and steel industry} \\
Particle size              & \SIrange{0.2}{600}{\upmu\meter}\\
\makecell[l]{Composition}         & \makecell[l]{$\approx 95\%$ Iron(III) oxide (\ce{Fe2O3})\\ $+\,5\%$ of other transition\\ metal oxides}\\\hline
\end{tabular}
\end{center}
\end{table}

\subsection{Experimental procedure}
Before the actual experiments with gas-powder two phase flow, the pressure drop curve of clean HTF passing through the clean packed bed depending on the HTF mass flux is recorded. For these experiments the HTF mass flux is set between \SIrange{0}{0.6}{\kilo\gram\per\square\meter\per\second}. The upper limit was chosen because a further increase of the HTF mass flux would increase the pressure drop and the thickness of the thermocline which both mean a low exergy efficiency of the PBTES. Empirical equations from the literature are utilized to reconstruct and validate the measured data. This procedure is repeated after every core experiment to record the pressure drop curve of the dusty packed bed.

For the core experiments of this study, the cell feeder is used to produce a gas-powder two phase flow with a relative powder content of \SIrange{0.025}{0.045}{\kilo\gram} powder per \SI{}{\kilo\gram} air which is representative for the powder content of exhaust gas from state-of-the-art steel producing processes (LD-converter and electric arc furnace). This gas-powder two phase flow further passes through the whole system as described in Section \ref{sec:setup}. The increase in pressure drop in the packed bed is measured with the differential pressure sensors PT1, PT2, ..., PT11 and the amount of powder hold-up inside the packed bed is determined by measuring the amount of powder that is feed into the HTF flow from the powder hopper and the amount of powder that is separated from the HTF in the powder filter after passing through the packed bed. When the pressure drop of the packed bed reaches a certain level/state of saturation the powder filter is cleaned and the flow direction of the HTF through the packed bed is reversed. Then clean air passes through the packed bed from the bottom to the top and exits the system through the powder filter. The reduction in pressure drop and powder hold-up is measured and determined as before. This whole procedure is repeated for a clean HTF mass flux of \qtylist[list-units = single]{0.3;0.4}{\kilo\gram\per\square\meter\per\second}.

\section{Theory and Calculations}
\label{sec:analysis}
For the validation of the measured data empirical equations from the literature are used. According to Kast et al. \cite{kast_l1_2013} there are two different ways to model the pressure drop in a fluid passing through a packed bed. The first and simpler option is, to model the packed bed as several pipes connected in parallel which leads to the \textit{Ergun}-equation \cite{ergun_fluid_1952}. With this equation the pressure drop per unit length can be calculated as

\begin{equation}
    \frac{\Delta p}{\Delta L} = 150\,\frac{\left(1-\psi\right)^2}{\psi^3}\,\frac{\eta\,v}{\overline{d}_{\mathrm{p}}^2} + 1.75\,\frac{1-\psi}{\psi^3}\,\frac{\rho_{\mathrm{f}}\,v^2}{\overline{d}_{\mathrm{p}}}
    \label{eq:ergun}
\end{equation}

where $\psi$ is the fractional void volume in the packed bed, $\overline{d}_{\mathrm{p}}$ is the Sauter-diameter of the packed bed material, $\eta$ is the dynamic viscosity of the fluid, $v$ is the superficial fluid velocity and $\rho_{\mathrm{f}}$ is the mass density of the fluid. At this point it should be mentioned, that \textit{Ergun}'s-equation is included in this study because it is the most common equation used to calculate the pressure drop in packed beds in the literature. However, one main disadvantage of the modeling approach used by \textit{Ergun} is that the real flow paths of the fluid are only insufficiently considered. Detailed information on the limitations of \textit{Ergun}'s-equation are provided by Kast et al. \cite{kast_l1_2013}.

A more versatile but also more complex equation for the calculation of the pressure drop per unit length is the \textit{Molerus}-equation. This equation was deduced based on the fluid flow around single particles and therefore allows a much more detailed modeling of the fluid flow paths within a packed bed with uniform randomly packed particles. By postulating equilibrium between the resistance force exerted by the fluid to each particle and the force due to the pressure drop in the fluid, Molerus \cite{molerus_koharente_1982} found that

\begin{equation}
    \frac{\Delta p}{\Delta L} = \frac{3}{4}\frac{\rho_{\mathrm{f}}\,v^2}{\overline{d}_{\mathrm{p}}}\,\frac{1-\psi}{\psi^2}\,\mathrm{Eu}\left(\Phi_{\mathrm{D}}\right)
    \label{eq:molerus}
\end{equation}

where 

\begin{equation}
    \begin{split}
    \mathrm{Eu}\left(\Phi_{\mathrm{D}}\right) = &\,\frac{24}{\mathrm{Re}\,\Phi_{\mathrm{D}}^2}\,\bigg\{1+0.685\,\bigg[ \frac{r_{\mathrm{0}}}{\delta_{\mathrm{0}}}+0.5\,\left(\frac{r_{\mathrm{0}}}{\delta_{\mathrm{0}}}\right)^2 \bigg]\,\bigg\} \\
    &\,+\frac{4}{\sqrt{\mathrm{Re}}\,\Phi_{\mathrm{D}}^{1.5}}\,\bigg[ 1+0.289\,\left(\frac{r_{\mathrm{0}}}{\delta_{\mathrm{0}}}\right)^{1.5} \bigg] \\
    &\,+\frac{1}{\Phi_{\mathrm{D}}}\,\bigg[ 0.4+0.514\,\frac{r_{\mathrm{0}}}{\delta_{\mathrm{0}}} \bigg] .
    \label{eq:euler}
    \end{split}
\end{equation}

The \textit{Euler}-number $\mathrm{Eu}\left(\Phi_{\mathrm{D}}\right)$ in Equation \eqref{eq:euler} is a function of the particle \textit{Reynolds}-number $\mathrm{Re}$, a factor $\Phi_{\mathrm{D}}$ that accounts for the non-spherical shape of the packed bed particles and a length ratio $r_{\mathrm{0}}/\delta_{\mathrm{0}}$ which is characteristic for the geometry of the fluid flow path between the packed bed particles. For a packed bed with a fractional void volume $\psi$ and uniform randomly packed particles the \textit{Reynolds}-number $Re$ and the length ratio $r_{\mathrm{0}}/\delta_{\mathrm{0}}$ can be calculated as

\begin{equation}
    \mathrm{Re} = \frac{\rho_{\mathrm{f}}\,v\,\overline{d}_{\mathrm{p}}}{\psi\,\eta}\,\,\,\,\,\,\,\,\mathrm{and}\,\,\,\,\,\,\,\,\frac{r_{\mathrm{0}}}{\delta_{\mathrm{0}}} = \bigg[ \frac{0.95}{\sqrt[3]{1-\psi}}-1 \bigg]^{-1} .
    \label{eq:reynolds}
\end{equation}

\subsection{Data processing and uncertainty analysis}
For a compact presentation of the most important findings of the present study the measured data from the experiments is processed using Equation \eqref{eq:dpnorm}. In this equation the pressure measurements from any two pressure measuring points $p_i$ and $p_j$ are used to compute the relative pressure drop between to measuring point $\Delta\hat{p}_{\mathrm{i-j}}(n)$. 

\begin{equation}
    \Delta\hat{p}_{\mathrm{i-j}}(n) = \frac{\Delta p_{\mathrm{i-j}}(n)}{\Delta p_{\mathrm{i-j}}(n = 1)} = \frac{p_i(n) - p_j(n)}{p_i(n = 1) - p_j(n = 1)}
    \label{eq:dpnorm}
\end{equation}

The relative pressure drop between two measuring points $\Delta\hat{p}_{\mathrm{i-j}}(n)$ is the ratio of the pressure difference of the $n^{\mathrm{th}}$ sample $\Delta p_{\mathrm{i-j}}(n)$ and the pressure difference of the first sample, i.e. the clean bed, $\Delta p_{\mathrm{i-j}}(n = 1)$.

To estimate the impact of uncertainties of the measurement devices on the results presented in this study an uncertainty analysis using the law of error propagation is conducted. As mentioned in Section \ref{sec:setup}, piezoresistive sensors with an accuracy of $\pm 0.06 \%$ of full scale are used to measure pressure differences between the measuring points in the test rig's packed bed. With a full scale of \SI{2000}{\pascal} (measurement range of $\pm \SI{1000}{\pascal}$) the utilized sensors deliver measurements with an accuracy of $\pm \SI{1.2}{\pascal}$. As the calculation of $\Delta\hat{p}_{\mathrm{i-j}}(n)$ in Equation \eqref{eq:error} requires two differential pressure measurements ($\Delta p_{\mathrm{i-j}}(n)$ and $\Delta p_{\mathrm{i-j}}(n = 1)$), the law of error propagation is used to calculate the uncertainty of the relative pressure drop as 

\begin{equation}
    \delta \Delta\hat{p}_{\mathrm{i-j}}(n)^2 = \left(\frac{\delta \Delta p_{\mathrm{i-j}}(n)}{\Delta p_{\mathrm{i-j}}(n = 1)}\right)^2 + \left(-\frac{\Delta p_{\mathrm{i-j}}(n)\,\delta \Delta p_{\mathrm{i-j}}(n = 1)}{\Delta p_{\mathrm{i-j}}(n = 1)^2}\right)^2.
    \label{eq:error}
\end{equation}

The relative uncertainty of $\Delta\hat{p}_{\mathrm{1-11}}(n)$ and $\Delta\hat{p}_{\mathrm{1-3}}(n)$ with respect to the calculated value are well below $\pm 2 \%$ and $\pm 8.5 \%$ respectively. These uncertainties are insignificant compared to the scattering of the experimental data (see Figures \ref{fig:poverj}, \ref{fig:powdermassflux} and \ref{fig:fluidmassflux}) and therefore do not have an impact on the quality of the results presented in Section \ref{sec:results}.

\section{Results and Discussion}
\label{sec:results}
\begin{figure}
    \centering
    \includegraphics[width=9cm]{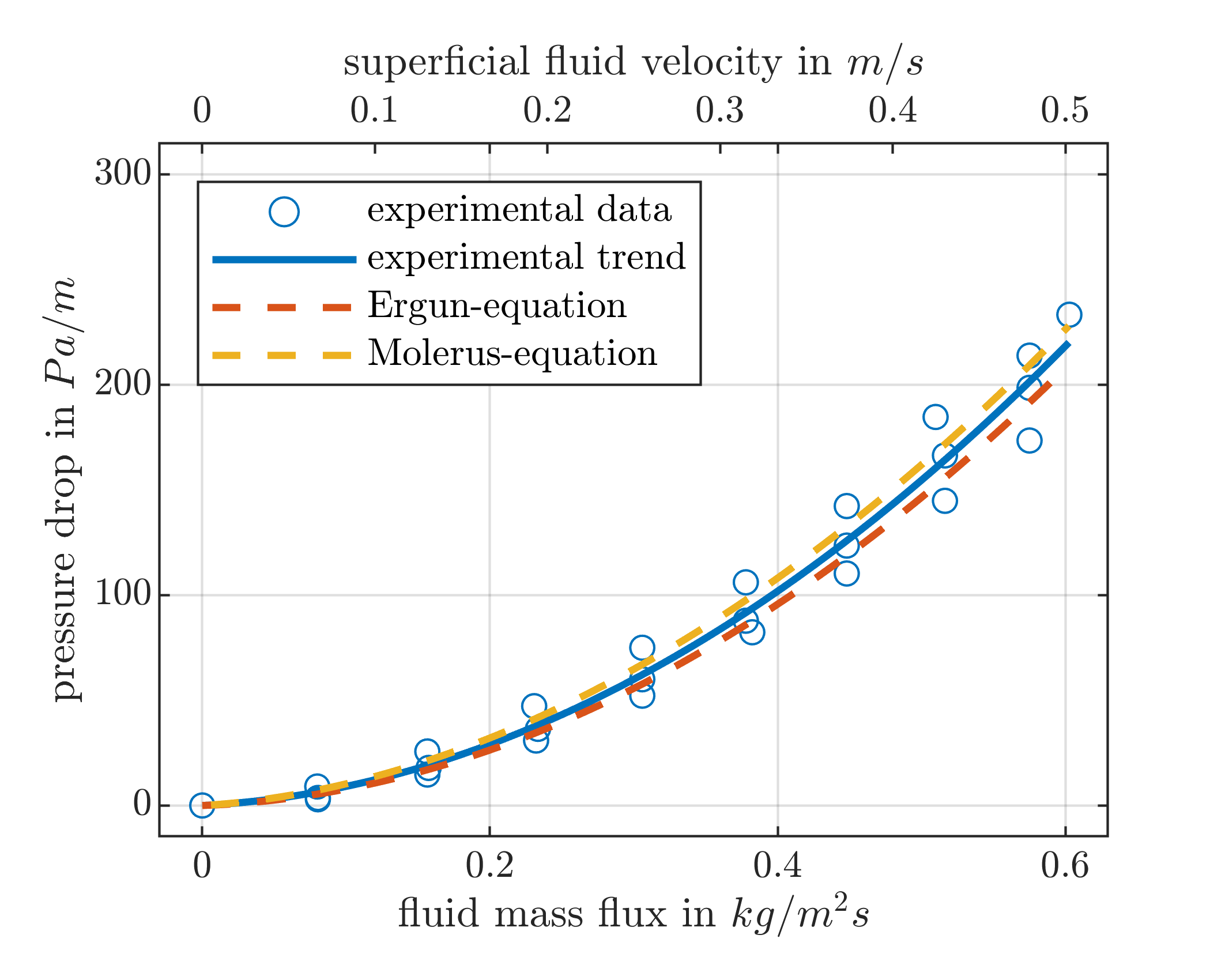}
    \caption{Pressure drop curve of clean HTF passing through a clean and dusty packed bed.}
    \label{fig:poverv_clean}
\end{figure}

To guarantee consistency and reproducibility of the measured data, the pressure drop curve of the clean packed bed is recorded multiple times before every core experiment. The results of these experiments are plotted in Figure \ref{fig:poverv_clean}. Furthermore, empirical equations from \textit{Ergun} and \textit{Molerus} are used to reproduce and validate the measured data. Both equations fit the data very well as it can be seen in Figure \ref{fig:poverv_clean}. Thereby it can be confirmed that the test rig used for the experiments in the present study delivers results that are consistent, reproducible and comparable with data from the literature.

\begin{figure}
    \centering
    \includegraphics[width=9cm]{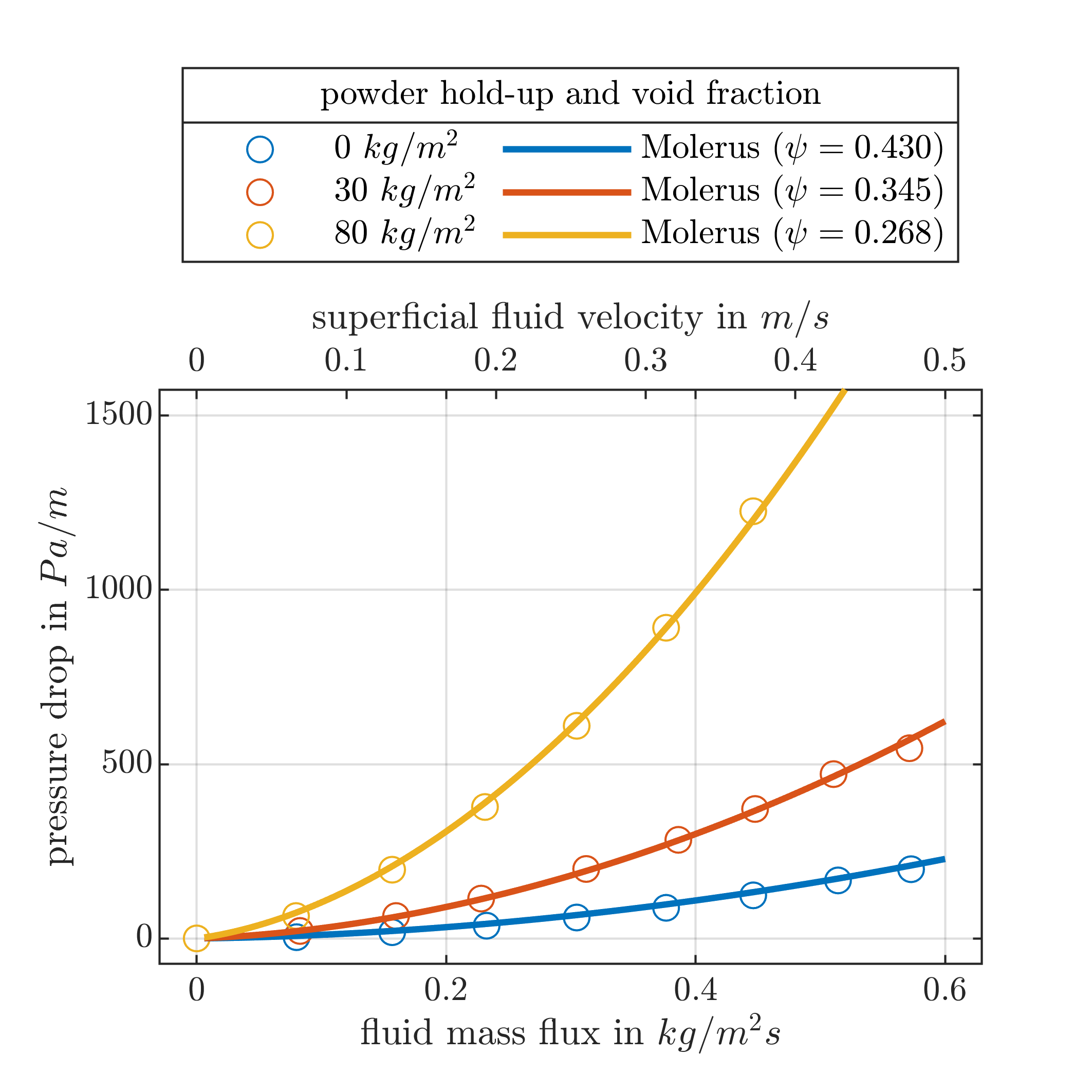}
    \caption{Impact of the powder hold-up on the pressure drop curve of a packed bed operated with clean HTF flow.}
    \label{fig:poverv}
\end{figure}

The impact of powder hold-up in the packed bed on the pressure drop curve can be seen in Figure \ref{fig:poverv}. In addition to the pressure drop curve of the clean packed bed this figure also shows the pressure drop curves of packed beds with a powder hold-up of $\qtylist[list-units = single]{30;80}{\kilo\gram\per\square\meter}$ (colored circles in Figure \ref{fig:poverv}). After adjusting/fitting the void fraction of the packed bed the empirical equation of \textit{Molerus} (solid lines in Figure \ref{fig:poverv}) still results in a very good fit of the measured data.

\subsection{Powder hold-up}
The amount of powder that accumulates in the packed bed during a charging process (powder hold-up) is determined by measuring the amount of powder that is fed into the system by the cell feeder and the amount of powder that accumulates in the powder filter. The experiments reveal, that for a fluid mass flux of $\SI{0.4}{\kilo\gram\per\square\meter\per\second}$ approximately 2\% of the powder is collected in the dust filter. Thus, 98\% of the powder accumulates in the packed bed during a charging process. This means, that the PBTES does not only act as a thermal storage, but also as a dust collector. The HTF that exits the PBTES during a charging process is not only cold but also carries just 2\% of the amount of powder than the HTF that enters the PBTES. Furthermore, discharging the storage with clean air that passes through the dusty packed bed in the opposite direction does not lead to a reduction of the powder hold-up in the packed bed. This is confirmed by a constant powder filter pressure drop observed during all the discharging phases. Hence, clean air that is used to discharge the PBTES is not contaminated with powder that has been accumulated in the packed bed in a previous charging process.

\begin{figure}
    \centering
    \includegraphics[width=9cm]{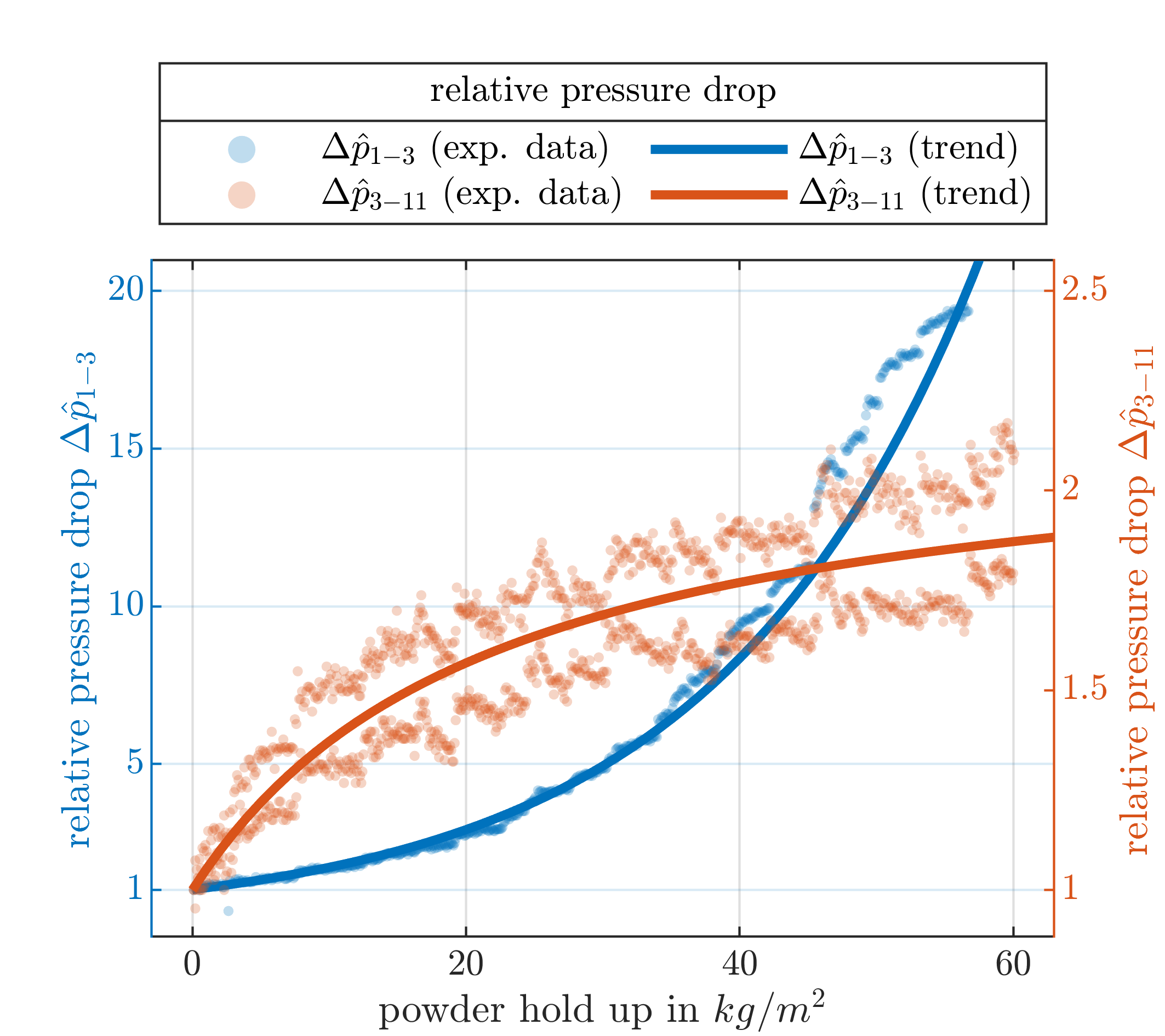}
    \caption{Vertical/axial distribution of the pressure drop in a packed bed with powder hold-up.}
    \label{fig:poverj}
\end{figure}

\begin{figure}
    \centering
    \includegraphics[width=9cm]{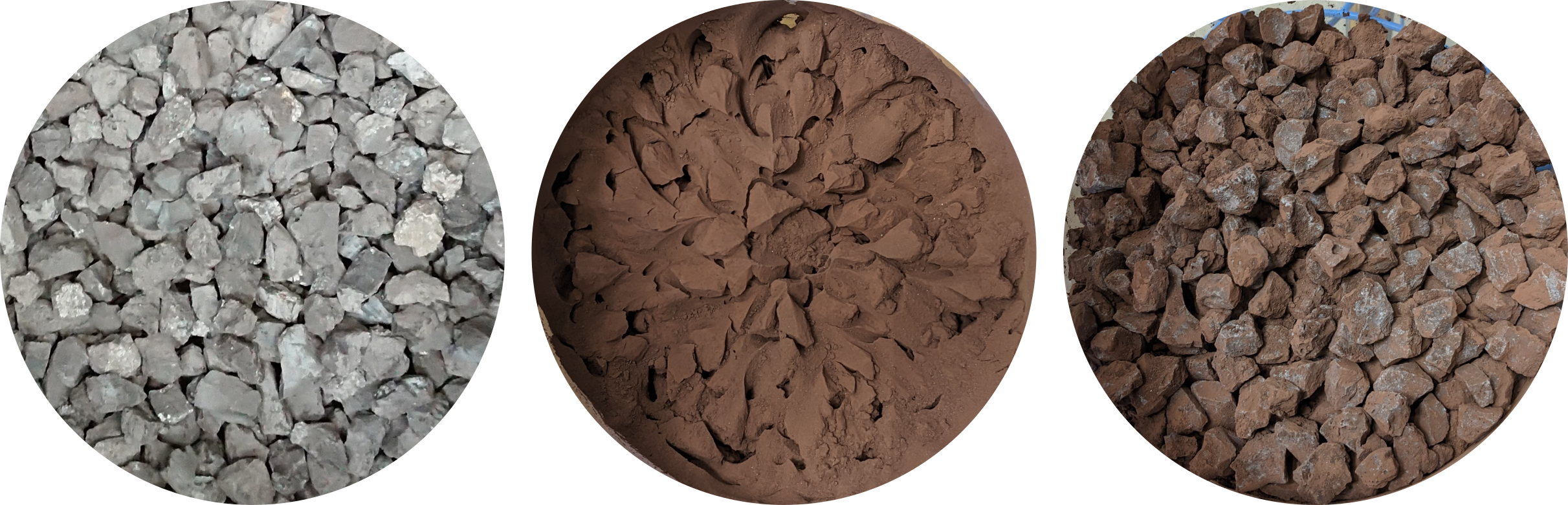}
    \caption{Photograph of a clean packed bed (left) and a packed bed with a powder hold-up of $\SI{30}{\kilo\gram\per\square\meter}$: top view directly after charging (center), top view after charging and after removing the upper fifth of the packed bed (right).}
    \label{fig:holdup}
\end{figure}

One of the most interesting observations from the experiments is the axial distribution of powder hold-up in the packed bed which is visualized in Figure \ref{fig:poverj}. This plot shows the relative pressure drop of the upper fifth of the packed bed -- where the gas-powder two phase flow enters the packed bed -- on the left ordinate ($\Delta p_{\mathrm{1-3}}$) and the relative pressure drop of the remaining height of the packed bed on the right ordinate ($\Delta p_{\mathrm{3-11}}$).The amount of powder hold-up in the packed bed is plotted on the abscissa. Both graphs start at a powder hold-up of $\SI{0}{\kilo\gram\per\square\meter}$ (clean bed) and increase with the powder hold-up in the packed bed. It can clearly be seen that the pressure drop in the upper fifth of the bed rises exponentially whereas the pressure drop in the lower region of the packed bed seems to flatten. At a powder hold-up of $\SI{40}{\kilo\gram\per\square\meter}$ the pressure drop in the upper fifth already increased by a factor of 8 while the pressure drop in the lower region of the packed bed has not even doubled. These results suggest, that a large part of the powder that is introduced into a PBTES accumulates near the surface at which the gas-powder two phase flow enter the packed bed. The small portion of the powder that is further transported into the system is evenly distributed inside the packed bed. These results are supported by examining the packed bed after a certain time of charging the test rig with a gas-powder two phase flow at the end of the experiments. Figure \ref{fig:holdup} shows the packed bed in the PBTES test rig at three different states. In the left picture a clean packed bed is depicted. The center picture shows the top surface (where the gas-powder two phase flow enters the packed bed) of a packed bed with a powder hold-up of $\SI{30}{\kilo\gram\per\square\meter}$. The right picture is taken from the same angle after the top fifth of the packed bed is removed. As was already assumed when interpreting the data in Figure \ref{fig:poverj}, Figure \ref{fig:holdup} shows that most of the powder hold-up in the packed bed is concentrated near the top surface of the packed bed. Furthermore the center photograph in Figure \ref{fig:poverj} shows an even radial distribution of the powder hold-up which indicates a uniform perfusion of the packed bed even with a significant amount of powder hold-up. It seems that there is no risk of random flow channel formation through the packed bed. 

\subsection{Influence of powder/fluid mass flux}
\begin{figure}
    \centering
    \includegraphics[width=9cm]{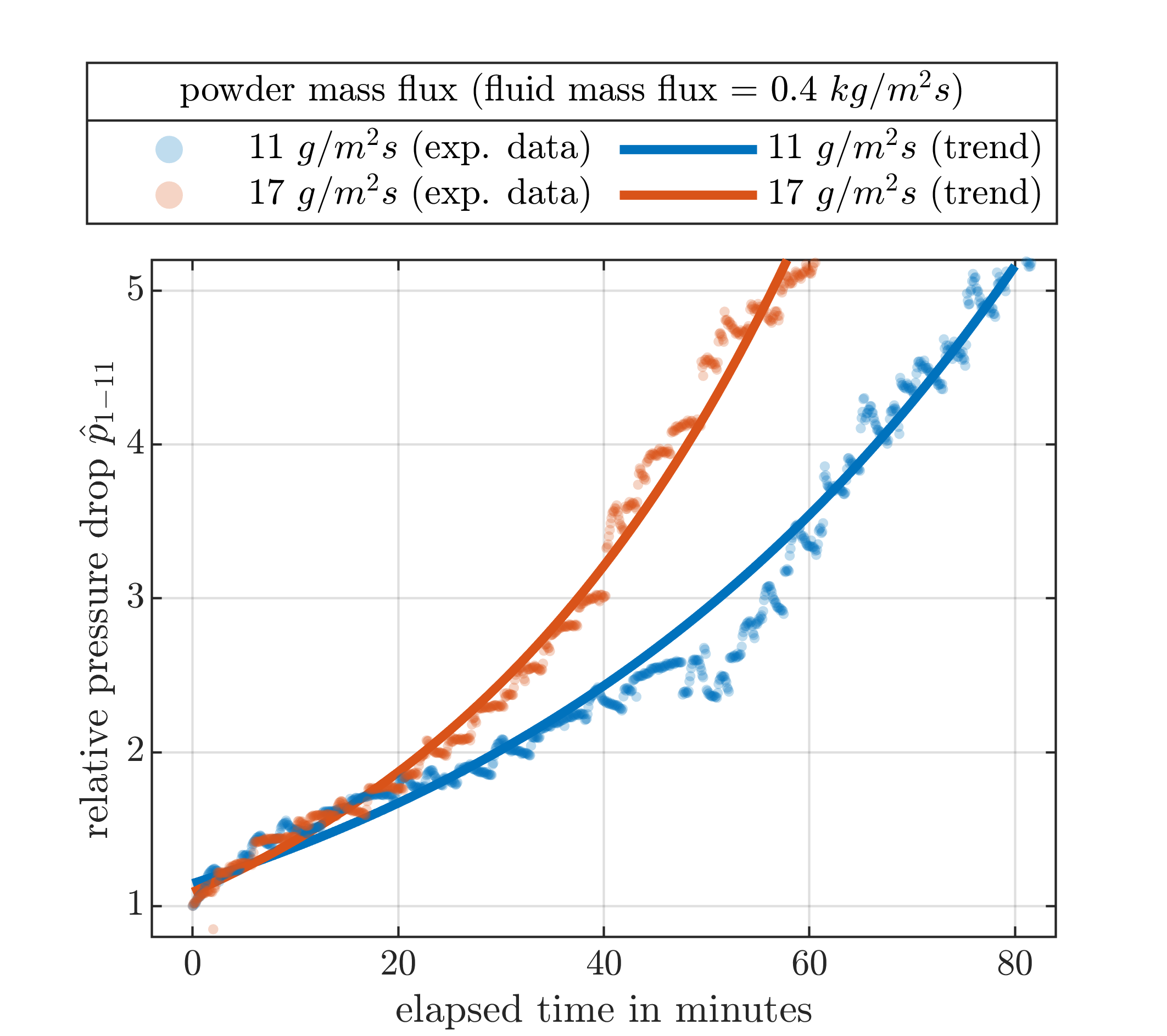}
    \caption{Impact of the powder mass flux on the pressure drop of a packed bed that is operated with a gas-powder two phase flow.}
    \label{fig:powdermassflux}
\end{figure}

The trend of the pressure drop for a charging period of \SI{80}{\minute} is depicted in Figure \ref{fig:powdermassflux}. This figure shows the relative pressure drop between the upper- and lowermost pressure measuring points $\Delta\hat{p}_{\mathrm{1-11}}$ of the test rig for two powder mass fluxes and a fluid mass flux of $\SI{0.4}{\kilo\gram\per\square\meter\per\second}$. These results show that the pressure drop of a PBTES that is charged with a gas-powder two phase exhaust gas increases exponentially over time. After a charging period of \SI{50}{\minute} with a powder mass flux of $\SI{17}{\kilo\gram\per\square\meter\per\second}$ an increase of the pressure drop by a factor of up to 4.5 is observed. A reduction of the powder mass flux also leads to a slower increase of the pressure drop, however, no saturation effects could be observed in any of the experiments. This means, that independent of the powder mass flux, powder will continue to accumulate over time which leads to a reduced exergy efficiency of the TES due to the increased pressure drop and eventually a clogging of the packed bed. Hence, frequent maintenance intervals at which the packed bed is cleaned or renewed are necessary. The frequency of maintenance can only be reduced by reducing the powder content of the fluid that passes through the packed bed. This could be realized by using a drop-out box that separates the coarse powder fractions from the HTF before it enters the PBTES.

\begin{figure}
    \centering
    \includegraphics[width=9cm]{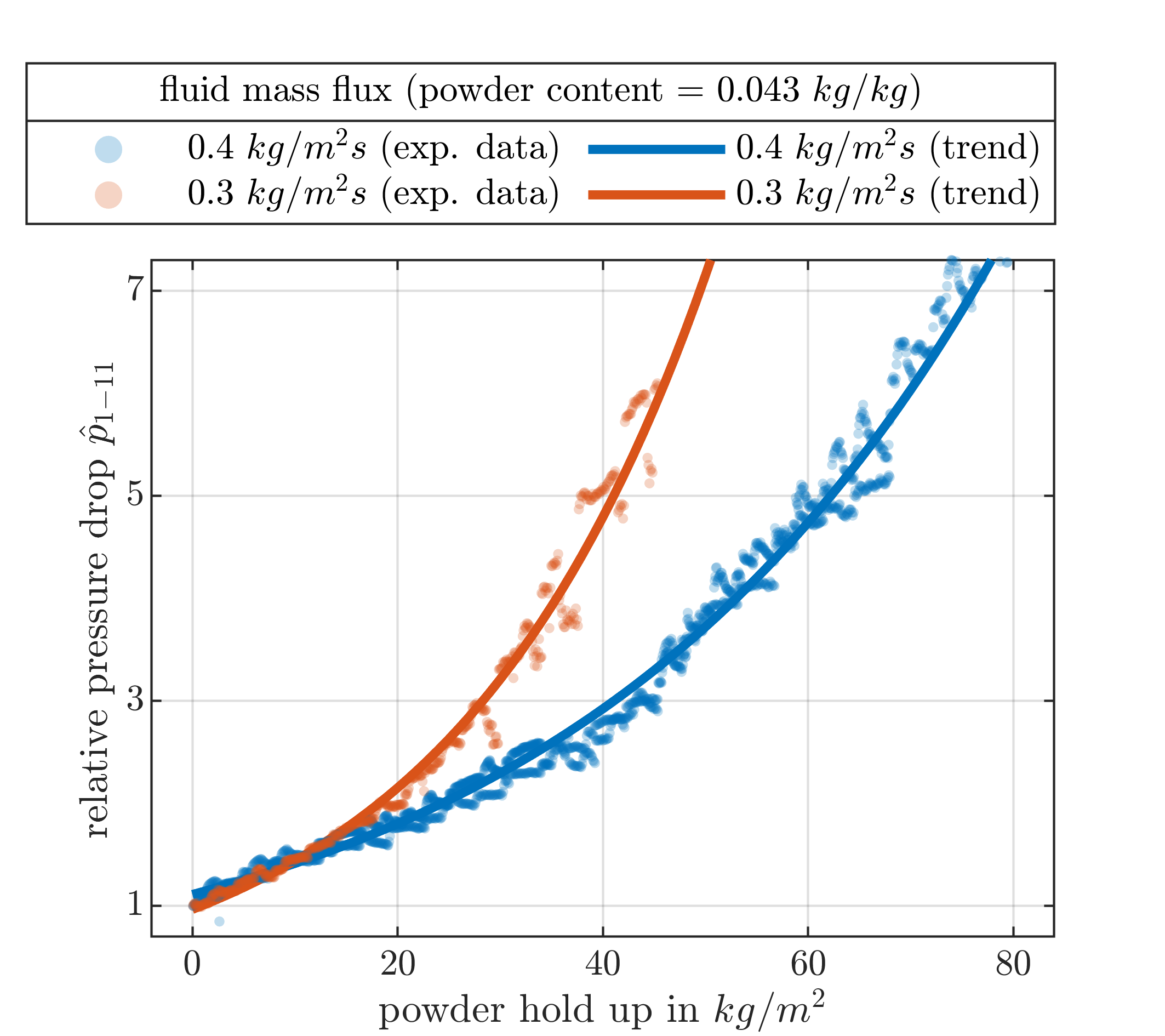}
    \caption{Impact of the fluid mass flux on the pressure drop of a packed bed that is operated with a gas-powder two phase flow.}
    \label{fig:fluidmassflux}
\end{figure}

In Figure \ref{fig:fluidmassflux} the impact of the fluid mass flux on the trend of the relative pressure drop of the packed bed is shown. It can be seen that a reduction of the fluid mass flux leads to a much faster increase of the relative pressure drop with respect to the powder hold-up in the packed bed. Notice, that in both cases the powder content of the fluid is $\SI{0.042}{\kilo\gram}$ powder per \SI{}{\kilo\gram} fluid and the abscissa in Figure \ref{fig:fluidmassflux} represents the powder hold-up in the packed bed and not the charging time. As a higher fluid mass flux with the same powder content also means a higher powder mass flux, the difference with respect to the charging time would not be as pronounced, but still present. The reason for that is based on the elutriation velocity of the powder particles which is (among other factors) mainly determined by its size. The smaller the particle, the lower its elutriation velocity and vice versa. For the present use-case this means that all particles with an elutriation velocity higher than the velocity of the fluid passing through the packed bed will accumulate in the packed bed. Particles with an elutriation velocity that is lower than the fluid velocity remain in the fluid flow and pass through the packed bed.
Nevertheless, the data in Figure \ref{fig:fluidmassflux} allows to make some important statements about the preferred operation strategy of a PBTES that is charged with a gas-powder two phase flow. Figure \ref{fig:fluidmassflux} shows that a powder hold-up of $\SI{40}{\kilo\gram\per\square\meter}$ increases the pressure drop of a PBTES that is charged with a fluid mass flux of $\qtylist[list-units = single]{0.3;0.4}{\kilo\gram\per\square\meter\per\second}$ by a factor of 5 and 3 respectively. This observation combined with the fact that both experiments were conducted with HTF having the same powder content means that the higher the HTF mass flux, i.e. the thermal power rate, at which the PBTES is charged, the slower the decrease of the storage's exergy efficiency due to the increased pressure drop. For the operation of a PBTES that is charged with a gas-powder two phase flow this implicates, that high charging power rates should be preferred to reduce the impact of powder hold-up on the pressure drop and hence the systems exergy efficiency. 

\subsection{Summary and Discussion}
The proposed integration of a PBTES for the waste heat recovery in the iron and steel industry is very promising due to its simplicity and its energetic and economic efficiency. The results presented in this section demonstrate that a PBTES that is operated as it is depicted in Figure \ref{fig:usecase} is capable of utilizing a considerable amount of the previously untapped waste heat potential in the iron and steel industry. Excess heat from steel producing processes that is available as hot gas-powder two phase exhaust gas can be stored in a PBTES by directly using the exhaust gas as HTF. Due to the even distribution of the powder hold-up in radial direction there is no risk of thermal performance degradation (storage capacity, thermal power rate) of the storage caused by random flow channel formation. As clean HTF that is used to discharge the PBTES is not contaminated by powder hold-up in the PBTES, heat recovered from the storage can be used in conventional waste heat boilers or for preheating purposes.

However, the results also show that there still is need for research and development work in order to make PBTES systems suitable for waste heat recovery in the iron and steel industry. The main challenge in this context is to find an operating/maintenance strategy to keep the pressure drop of HTF passing through the packed bed within appropriate bounds. It is important to keep the pressure drop of HTF passing through the packed bed as low as possible as the exergy efficiency of a PBTES is strongly influenced by the energy needed to pump HTF through the packed bed. Considering the results presented in this study strategies and technologies to reduction of the powder content of the HTF before it enters the PBTES and to reduce the amount of powder hold-up that accumulates in the PBTES are required. The powder content of the HTF can be reduce with existing technologies like gravity separators (drop-out boxes). An interesting approach to reduce the powder hold-up in the packed bed is to switch the flow direction for the charging and discharging process of the PBTES (charging from bottom and discharging from top) and to utilize periodic knocking/trembling mechanisms to clean the packed bed from the accumulated powder hold-up. As the powder hold-up in the PBTES concentrates near the surface at which the HTF enters the packed bed, which, for switched flow directions would be the bottom surface of the bed, the removal of the powder hold-up takes place by gravitation. The limitations and possible challenges this approach could entail include a reduced exergy efficiency of the storage due to thermocline degradation especially during long standby periods. For details considering these effects the authors refer to their previous work \cite{schwarzmayr_standby_2023}.

\section{Conclusion}
\label{sec:conclusion}
The present study examined the suitability of packed bed thermal energy storage systems for the waste heat recovery in the iron and steel industry. Besides extreme temperatures and the highly fluctuating and unpredictable availability of excess heat, the main difficulty of waste heat recovery in industrial processes is the high amount of powder that is transported by the hot exhaust gases. Therefore the focus of this study was the investigation of the behaviour/characteristics of a packed bed thermal energy storage that is operated with a gas-powder two phase heat transfer fluid.

A lab-scale test rig of a packed bed thermal energy storage was used to quantify the pressure drop and powder hold-up that have to be expected when a packed bed thermal energy storage is operated with a gas-powder two phase heat transfer fluid. 

The investigations revealed that, for the given materials and operational parameters, 98 \% of the powder that enters the test rig when it is charged with a gas-powder two phase fluid accumulates in the packed bed. Only the finest fractions (2 \%) of the powder remain in the fluid flow passing through the packed bed. Furthermore, reversing the flow direction of the heat transfer fluid and discharging the test rig with a clean single phase fluid does not lead to a reduction of the powder hold-up inside the packed bed. Hence, clean fluid that is used to discharge the test rig is not contaminated with the powder hold-up in the packed bed. Additionally an uneven distribution of the powder hold-up along the fluid flow direction was observed by examining the packed bed after the experiments. These observations are consistent with the pressure drop measurements during the experiments. A large proportion of the powder hold-up is concentrated near the surface at which the fluid flow enters the packed bed. The distribution of powder hold-up in radial direction of the test rig was uniform and no random flow channel formation could be observed.

Overall, the integration of a packed bed thermal energy storage as waste heat recovery system in the iron and steel industry was found to be very promising and is definitely worth further investigation. Especially the evaluation and development of suitable strategies for the removal of powder hold-up from the storage material should be the main focus in future research projects.

\section*{Acknowledgement}
The authors acknowledge funding support of this work through the research project \textit{5DIndustrialTwin} as part of the Austrian Climate and Energy Fund's initiative Energieforschung (e!MISSION) 6\textsuperscript{th} call (KLIEN/FFG project number 881140). Furthermore, the authors acknowledge TU Wien Bibliothek for financial support through its Open Access Funding programme.

\section*{Declaration of Competing Interest}
The authors declare that they have no known competing financial interests or personal relationships that could have appeared to influence the work reported in this paper.

\appendix


 \bibliographystyle{elsarticle-num} 
 \bibliography{main.bib}


\printnomenclature
\nomAcro[tes]{TES}{Thermal energy storage}
\nomAcro[pbtes]{PBTES}{Packed bed thermal energy storage}
\nomAcro[htf]{HTF}{Heat transfer fluid}
\nomAcro[ld]{LD}{Linz-Donawitz}
\nomAcro[eaf]{EAF}{Electric arc furnace}
\nomAcro[csp]{CSP}{Concentrated solar power}

\nomRoman[dover]{$\overline{d}$}{Sauter diameter in \si{\meter}}
\nomRoman[Eu]{$\mathrm{Eu}$}{\textit{Euler}-number}
\nomRoman[L]{$L$}{Packed bed height in \si{\meter}}
\nomRoman[p]{$p$}{Pressure in \si{\pascal}}
\nomRoman[p]{$\Delta p$}{Pressure drop in \si{\pascal}}
\nomRoman[p]{$\Delta\hat{p}$}{Relative pressure drop}
\nomRoman[r0]{$r_{\mathrm{0}}$}{Characteristic length of a non-spherical particle in \si{\meter}}
\nomRoman[Re]{$\mathrm{Re}$}{\textit{Reynolds}-number}
\nomRoman[v]{$v$}{Superficial fluid velocity in \si{\meter\per\second}}

\nomGreek[delta]{$\delta_{\mathrm{0}}$}{Characteristic length for the fluid flow path in \si{\meter}}
\nomGreek[psi]{$\psi$}{Fractional void volume in \si{\cubic\meter\per\cubic\meter}}
\nomGreek[PHI]{$\Phi_{\mathrm{D}}$}{Shape factor for non-spherical particles}
\nomGreek[n]{$\eta$}{Dynamic viscosity of the fluid in \si{\pascal\second}}
\nomGreek[r]{$\rho$}{Mass density of the fluid in \si{\kilo\gram\per\cubic\meter}}

\nomSub[i]{$i$}{Index of pressure measuring point}
\nomSub[j]{$j$}{Index of pressure measuring point}
\nomSub[n]{$n$}{Sample number}
\nomSub[f]{$f$}{Fluid}
\nomSub[p]{$p$}{(packed bed) Particle}





\end{document}